\documentclass[aps,amsmath,amssymb,superscriptaddress,nofootinbib,twocolumn,floatfix]{revtex4-1}

\usepackage{graphicx}
\usepackage[usenames]{color}
\usepackage[colorlinks=true,linkcolor=cyan,citecolor=cyan,anchorcolor=green,urlcolor=magenta]{hyperref}
\usepackage{float}
\usepackage{subfigure}
\usepackage{ulem}

\def\be{\begin{equation}}
\def\ee{\end{equation}}
\def\bea{\begin{eqnarray}}
\def\eea{\end{eqnarray}}

\newcommand{\ket}[1]{\mbox{$| #1 \rangle$}}

\begin{document}

\title{Spontaneous magnetism after a ``remote" quench: a proposal to test Copenhagen interpretation}
\author{Jianda Wu}
\email{wujd@sjtu.edu.cn}
\affiliation{Tsung-Dao Lee Institute \& School of Physics and Astronomy, Shanghai Jiao Tong University, Shanghai, 200240, China}

\author{Wenxin Ding}
\email{wxding@ahu.edu.cn}
\affiliation{School of Physics and Optoelectric Engineering, Anhui University, Hefei, Anhui Province, 230601, China}

\date{\today}

\begin{abstract}
A change of quantum states for a quantum particle may lead to a change of physical
field it exerts to the environment. We discuss such Gedankenexperiment for measuring the magnetic dipole fields associated with the electronic spins. 
When entangled,  electrons are no longer free Dirac electrons but become an excited state in a quantum-electrodynamics sense. A measurement of magnetism associated with entanglement-collapse of electrons becomes a test for Copenhagen interpretation in the reign of quantum-electrodynamics. This proposal is equally applicable to other particles and their associated fields and interactions, such as neutrons and the electroweak force.
\end{abstract}

\maketitle
\textbf{Introduction.}
Quantum information, such as spin state or other quantized information,
is carried on by physical quantum particles or quantum devices.
A change of quantum state may lead
to a change of physical field the particle exerts to environment.
For example when an electron is in a singlet state
quantum entangled with another electron [Eq.~(\ref{qc})], the electron is in an unpolarized state
with absence of magnetic field generation. When the electron ``collapses"
to a definitive  polarized state, it has to restore the magnetic field surrounding it which is tied to its dipole moment.
On the other hand,from Copenhagen interpretation (CI), a measurement leads to a collapse of a quantum state which ``instantaneously" transits to certain measurement basis with certain probability.
Furthermore, when applying CI to the measurement of
a two-body quantum entangled state, once one body is measured and
collapses to a certain state,
the other body, no matter how far they are separated,
would immediately and correlatedly collapse to a definitive
state.
Here, we show that via the field change due to a state change of a quantum
particle, CI on collapse and correlated collapse scheme for measurement
of a quantum entangled state can in principal be tested. In the remainder,
we first describe the details of the proposal, then come
back to discuss various possible outcomes of the proposed experiments.

\textbf{Observation:
the collapse of spin-entangled electrons leads to a quench of local magnetic dipole moment.}

Suppose Alice and Bob share a quantum channel of a spin-1/2
singlet state composed by two electrons
(or other quantum particles/devices carrying spin 1/2):
\be
\left| \psi  \right\rangle   = \frac{1}{{\sqrt 2 }}\left( {\left|  \uparrow  \right\rangle _{A} \left|  \downarrow  \right\rangle _{B}  - \left|  \downarrow  \right\rangle _{A} \left|  \uparrow  \right\rangle _{B} } \right),\label{qc}
\ee
where $A$ and $B$ are used to denote Alice and Bob
, and ${\left\{ {\left|  \uparrow  \right\rangle ,\left|  \downarrow  \right\rangle } \right\}_{A\;{\rm{or}}\;B}}$
are eigenstates of Pauli matrix $\sigma_z^{A \;{\rm or}\; B}$ with eigenvalues $\pm 1$.
Before Alice carries out measurement,
Bob's electron stays in an unpolarized state
along any direction thus without magnetic field
surrounded it. After Alice' polarization projection measurement (Here
assuming Alice carried out polarization projection measurement along $z$ direction,
which is also known for Bob), Alice may obtain her electron either along
$+z$ or $-z$ direction. Following CI measurement scheme for the state of
Eq.~(\ref{qc}), Bob's electron
would immediately collapse to a state with a definitive polarized state
either along $-z$ or $+z$ direction, respectively. No matter which
polarized state Bob's electron collapses to, Bob's electron will
gain a finite magnetic moment at the moment it collapses, and begins
to generate an electron-magnetic (EM) field that propagates in the space,
and eventually relaxes to a stable magnetic dipole field leading to
a finite magnetism just as a free electron. Such a change of magnetism for Bob's electron
due to the ``remote" quench of Alice' electron should be detectable by
a proper magnetic-flux experimental setup.

Semi-classically or from a non-relativistic quantum mechanical perspective, a polarized (or free) electronic spin possesses both an electric field and a magnetic dipole field. While the electric field remains unchanged, the dipole field is directly determined by the electronic spin state. Therefore, the collapse of an entangled spin-singlet state is inevitably accompanied by a quench of the local magnetic dipole field.  Since direct measurement of the free electron's dipole field should be equivalent to measurement of the electronic spin, the standard quantum information results are still valid.
Although the collapse of the spin state is measurably instantaneous, the quench of the dipole field is a {\bf dynamical} process: the dipole field cannot be restored instantaneous from a semi-classical point of view but rather can only propagate and be restored at the speed of light.
We must also note that the dipole field possesses a non-zero (classically divergent) energy. Therefore, it is fundamentally a quantum-electromagnetic-dynamics (QED) process.

In relativistic quantum theory, a pair of spin-singlet entangled electrons is an excited state comparing to two free electrons.
For a single Dirac electron, its spin is not longer a good quantum number. Instead, with an electron's spin $\boldsymbol S$ and momentum $\boldsymbol p$, the electronic helicity
$\lambda_s = {\boldsymbol S} \cdot {\boldsymbol p} / |{\boldsymbol p}|$ commutes with the Dirac Hamiltonian,
whose eigenvalues $\lambda_{S,\pm} = \pm 1/2 \hbar$ serve as good quantum numbers. When ${\boldsymbol p_1} \parallel {\boldsymbol p_2} $,
the helicity singlet $|hs\rangle = 2^{-1/2}(|\lambda_S=1/2, {\boldsymbol p_1}\rangle \otimes |\lambda_S=-1/2, {\boldsymbol p_2}\rangle - |\lambda_S=-1/2, {\boldsymbol p_1}\rangle \otimes |\lambda_S=-1/2, {\boldsymbol p_2}\rangle )$ is then also a spin singlet.
Taking $p_1 \sim p_2 \rightarrow 0$ limit, we recover the
non-relativistic limit of a pure spin singlet state.
For ${\boldsymbol p_1} \nparallel {\boldsymbol p_2} $, even the helicity singlet is no longer stationary.
However, helicity is not invariant under a Lorentzian boost. On the other hand,
the spin and chirality is invariant under a Lorentzian boost but not preserved for a free electron.
In fact, we can view the preparation of a two-electron entangled state from two free-electrons as a nontrivial scattering process. The semi-classical dipole field energy can also be understood as the electron's self-energy due to self-interaction via virtual photons.

Consequently, the process of measurement induced collapse should be viewed as a QED process.
When the local state collapses, the electronic self-energy also experiences a quench.
The difference in electronic self-energy will be transferred and turned into the quench of the dipole field, or emission of photons.

The above process does not violate energy conservation locally.
Because when an electron is quantum entangled with the other electron in the from of
Eq.~(\ref{qc}), it is not longer an eigenstate of Dirac Hamiltonian for
the free electron but, instead, in an excited state. This implies
in the infra-red limit an
electron in quantum entangled state will have a
larger self energy $M_1$ compared that of
an independent electron $M_0$. When Bob's electron collapses from a quantum entangled state
to an independent one, the electron transits from higher self energy state
to self energy in ground state. Such energy difference
provides the necessary energy to generate a photon for the system.

The measurement of the emitted photon cannot teleport information, which is consistent with standard quantum information theory.
Consider the angular momenta conservation of the photon emission process. Locally, the angular momenta change of the electronic spin collapse is 1/2, which means locally a photon emission is forbidden. However, globally, i.e. for the two-electron entangled state, the constraint of total $\Delta L$ being an integer can be satisfied. For simplicity, we consider $\Delta L = 0$ as an example. A permissible state of the emitted photon is a spatial superposition state such as $(\ket{A:R} + \ket{B:L})/\sqrt{2}$. Then, Alice has a $P=1/2$ chance of measuring the photon, hence no information is teleported. However, a measurement of the emitted photon will be
a fundamental test on QED.


\textbf{The proposals.}
Our observation opens up a new playground for testing quantum mechanics,
quantum information and even quantum entanglement in QED. Below, we propose and discuss two experimental measurements concerning two different aspects of this collapse-induced quench process: i) the dynamics of the quench, i.e. the photon-emission following the quantum collapse, and ii) measurement of the magnetic dipole moment after the quench.

{\it Proposal I: measurement of the collapse-induced photon emission.}
As our analysis in previous section shows, a pair of spin-entangled electrons is not a simple product state of free Dirac electrons, but rather an excited state from the QED perspective. Collapse of the state leads to photon emissions and reduction of the electrons' self-energy. Effects of path-entanglement on electronic Cherenkov photon emission\cite{karnieli} have been studied. We propose that both theoretical and experimental investigation of the photon emission of spin-entangled electrons could lead to new phenomena as well as providing new information on entanglement properties in QED.

More importantly, the observation and analysis of this work is not restricted to electrons and QED, but equally applicable to other spinful particles with magnetic dipole moments or even electric dipole moments. Taking neutrons, which only have magnetic dipole moments, as an example. The collapse-induced quench process of a pair of spin-entangled neutrons should also lead to photon emission, but for its lack of electric charge, the process is likely dominated by electroweak interactions.

{\it Proposal II: measurement of magnetic flux generated from the magnetic dipole moments.}
After the photon emission process, the spin-associated magnetic dipoles should also be measurable in principle.
Here, we propose that it is possible for Bob to detect the collapse through {\it magnetic flux measurements}, such as via a nanoSQUID\cite{nanoSQUID}, with or even without classical communication.
Without classical communication, the magnteic flux would appear when Bob's electron
collapses after Alice' measurement.

The flux measurements provide a route to test CI about its interpretation on collapse and
correlated collapse of wavefunction measurement.
Measuring the magnetic flux provides
an evidence to help Bob justify whether Alice has
carried out measurement.
With a preexisting protocol set up via classical
communication, Bob can distinguish whether Alice has made a measurement or not.
Possible loopholes about this scenario are further discussed in the end of the last section.


In real experimental setups for both proposals, there is no restriction to choose
electrons to prepare the spin entangled state. One can also choose
neutral atoms or quantum device,
such as $^3\text {He}$, neutron, Nitrogen-Vacancy center in diamond,
or quantum dot, etc..
 The polarized spin state after quench can
provide a magnetic flux ${\Phi}(a) = \frac{{{\mu _0}}}{2}\frac{{{\mu _B}}}{a}$
with vacuum permeability $\mu_0$ and Bohr magneton $\mu_B$; and $a$
is the radius of a circle area that encloses the electron. If choose $a$
at the scale of $0.1 \mu m$, then the generated magnetic flux is at the order of
$10^{-22} \text{Wb}$ which is detectable via a nanoSQUID\cite{nanoSQUID}.
Experimental difficulty could be further reduced by considering
a GHZ (Greenberger-Horne-Zeilinger) state \cite{GHZ},
$\left| \psi  \right\rangle _{\text{GHZ}}  = \frac{1}{{\sqrt 2 }}\left( {{{\left| { \uparrow   \cdots  \uparrow } \right\rangle }_A}{{\left| { \downarrow   \cdots  \downarrow } \right\rangle }_B} - {{\left| { \downarrow   \cdots  \downarrow } \right\rangle }_A}{{\left| { \uparrow    \cdots  \uparrow } \right\rangle }_B}} \right)$
where Alice and Bob share quantum entangled $2N$ spins with $N$ spins
at each place. A GHZ state can in principle
enhance the magnetic flux by $N$ times after
it collapses.

\textbf{Discussion.}
The information from different consequences of measurement versus
no measurement does not
spoil non-signaling theorem or no-communication theorem\cite{peres,popescu,florig} in quantum information theory, because Alice and Bob need to share in advance one bit of information on the polarization
of spin that Alice measures. Nevertheless, a realization of our proposal can
guarantee the information transmission becomes absolutely secure without
encryption coding.

The key for the proposal is based on change of surrounding
physical field due to change of quantum state carried on
by quantum particles.
As such, similar setups are not limited to the observation of magnetic-field but
also the change of other physical fields
associated with different fundamental forces.


The theoretical observation and the associated experimental proposals of this work present new challenges for QED and quantum field theories in general. If one takes a local perspective, one needs to compute processes with a mixed density matrix as the initial state, which is not available in QED to the best of the authors' knowledge. If one takes a global perspective, the separation between the two electrons can be space-like. QED computation for a space-like initial state is also a new challenge. Moreover, our proposals can serve as experimental tests for relativistic causality of quantum operations\cite{beckman}.

Finally, we discuss possible loopholes.
The emitted photons can be measured far away from the particles, thus the measurement is unlikely to
interfere with the entangled state before Alice makes a measurement. However, a flux measurement typically
requires a preexisting device, a nanoSQUID for example, in the vicinity of the particle.
Though the environment of a nanoSQUID could be extremely clean \cite{nanoSQUID},
the possibility that even an ideal device would interact with thus
disentangle the entangled particles before Alice' measurement
could not be excluded.
We are looking forward to a careful experiment to
test our proposals.

\textbf{Acknowledgement.}
We thank helpful discussion with Yongde Zhang, Thors Hans Hansson,
Wei Ku, Qingdong Jiang, Chi-Ming Yim, Rong Yu, Jie Ma, and Yu-Ao Chen. J.W. thanks support
by the Innovation Program for Quantum Science and Technology No. 2021ZD0301900, the Natural
Science Foundation of Shanghai with grant No. 20ZR1428400, and Shanghai Pujiang
Program with grant No. 20PJ1408100. J.W. acknowledges additional support from a Shanghai
talent program.
W.D. thanks support by the Anhui Provincial Natural Science Foundation Young Scientist
Grant number 1908085QA35 and the Startup Grant number S020118002/002 of Anhui University.

\end{document}